\documentclass{article}
\usepackage{spconf,amsmath,graphicx}
\usepackage{booktabs}
\usepackage{graphicx}
\usepackage{float}
\usepackage{enumitem}
\setlist{nosep, leftmargin=14pt}
\usepackage{hyperref} 

\usepackage{mwe} 

\title{EViT-Unet: U-Net Like Efficient Vision Transformer for Medical Image Segmentation on Mobile and Edge Devices}
\name{Xin Li$^{1}$, Wenhui Zhu$^{1}$, Xuanzhao Dong$^{1}$, Oana M. Dumitrascu$^{2}$, Yalin Wang$^{1}$}

\address{$^{1}$ School of Computing and Augmented Intelligence, Arizona State University, AZ 85281, USA
\\
$^{2}$ Department of Neurology, Mayo Clinic, Scottsdale, AZ 85251, USA
}
\begin{document}
%
\maketitle
\begin{abstract}
With the rapid development of deep learning, CNN-based U-shaped networks have succeeded in medical image segmentation and are widely applied for various tasks. However, their limitations in capturing global features hinder their performance in complex segmentation tasks. The rise of Vision Transformer (ViT) has effectively compensated for this deficiency of CNNs and promoted the application of ViT-based U-networks in medical image segmentation. However, the high computational demands of ViT make it unsuitable for many medical devices and mobile platforms with limited resources, restricting its deployment on resource-constrained and edge devices. To address this, we propose EViT-UNet, an efficient ViT-based segmentation network that reduces computational complexity while maintaining accuracy, making it ideal for resource-constrained medical devices. EViT-UNet is built on a U-shaped architecture, comprising an encoder, decoder, bottleneck layer, and skip connections, combining convolutional operations with self-attention mechanisms to optimize efficiency. Experimental results demonstrate that EViT-UNet achieves high accuracy in medical image segmentation while significantly reducing computational complexity.The code is available at \url{https://github.com/Retinal-Research/EVIT-UNET}.
\end{abstract}
%


\begin{figure}[htb]

  \centering
  \centerline{\includegraphics[width=8cm]{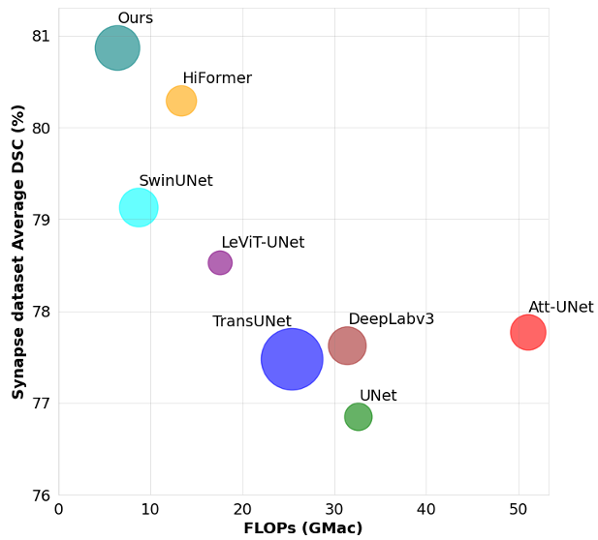}}
%
\caption{Comparison of model size, FLOPs, and performance (Dice on Synapse dataset). The area of each circle is proportional to the number of parameters (model size). }
\label{fig:Flops}
\end{figure}

\section{Introduction}
\label{sec:intro}
With the rapid development of deep learning, the field of medical image analysis has made significant progress, especially in image segmentation tasks. U-shape networks have become the mainstream model for segmentation networks. The structure integrates a symmetric encoder-decoder configuration with a bottleneck layer and skip-connection, constituting the classic U-shaped segmentation network~\cite{U-net}. The classical U-Net architecture employs convolutional downsampling to capture multi-level features. It incorporates skip connections to preserve spatial details, followed by decoder-based upsampling to reconstruct image resolution for precise pixel-level segmentation. This structure has succeeded greatly in various medical image segmentation tasks, such as heart, organ, and lesion segmentation. Some works based on U-Net aim to enhance network performance, such as U-Net++~\cite{unet++}, which improves network efficiency by refining the skip connection mechanism. Furthermore, DeepLabV3~\cite{deeplabv3} enhances the processing of multi-scale features by optimizing convolutional operations. However, CNN-based UNet still needs to improve in capturing global semantic information and handling complex feature interactions.

The introduction of Vision Transformers (ViTs) addresses the limitations of CNNs in capturing global information. The self-attention mechanism in ViTs effectively captures global context~\cite{attention}, providing significant advantages over traditional CNNs in managing global features and long-range dependencies. Researchers have started exploring their application in medical image segmentation. For instance, Att-UNet~\cite{att-Unet} and TransUNet~\cite{transunet} introduced Transformers to the UNet network, and hybrid CNN-Transformer models like HiFormer~\cite{hiformer} and UCTransUNet~\cite{uctransnet} were designed to improve network speed. Furthermore, SwinUNet~\cite{swin-Unet} and MedT~\cite{medt} employ specialized Transformer architectures to enhance network speed and performance. The self-attention mechanism in ViTs improves segmentation accuracy and robustness, further advancing effectiveness in medical image segmentation tasks.

Although ViTs exceptionally perform in vision tasks, their high computational complexity limits their applicability on resource-constrained devices~\cite{lightweightViT,unext}. In medical image segmentation tasks, reducing computational complexity while maintaining accuracy has always been a key pursuit in segmentation tasks for resource-constrained devices.~\cite{swin-Unet, hiformer} Thus, we propose EViT-UNet, an efficient U-shaped network based on ViT for medical image segmentation on mobile and edge devices. It inherits ViT’s ability to capture global information while reducing computational complexity through the combination of convolution and self-attention mechanisms, ensuring high accuracy while minimizing computational cost, making it ideal for mobile and edge devices. Tested on multiple datasets, EViT-UNet demonstrated superior segmentation accuracy and outperformed other popular segmentation frameworks. Our key contributions are as follows: (1) Developing an efficient U-shaped segmentation framework based on ViT that integrates an encoder, a decoder, and skip connections, which has shown outstanding performance across diverse datasets. (2) Achieving the best computational efficiency in comparative analyses with multiple networks. (3) Successfully reducing computational complexity while maintaining high accuracy, enhancing the feasibility of deploying this technology in resource-limited settings for medical image segmentation tasks.

\begin{figure}[!ht]

  \centering
  \centerline{\includegraphics[width=8.5cm]{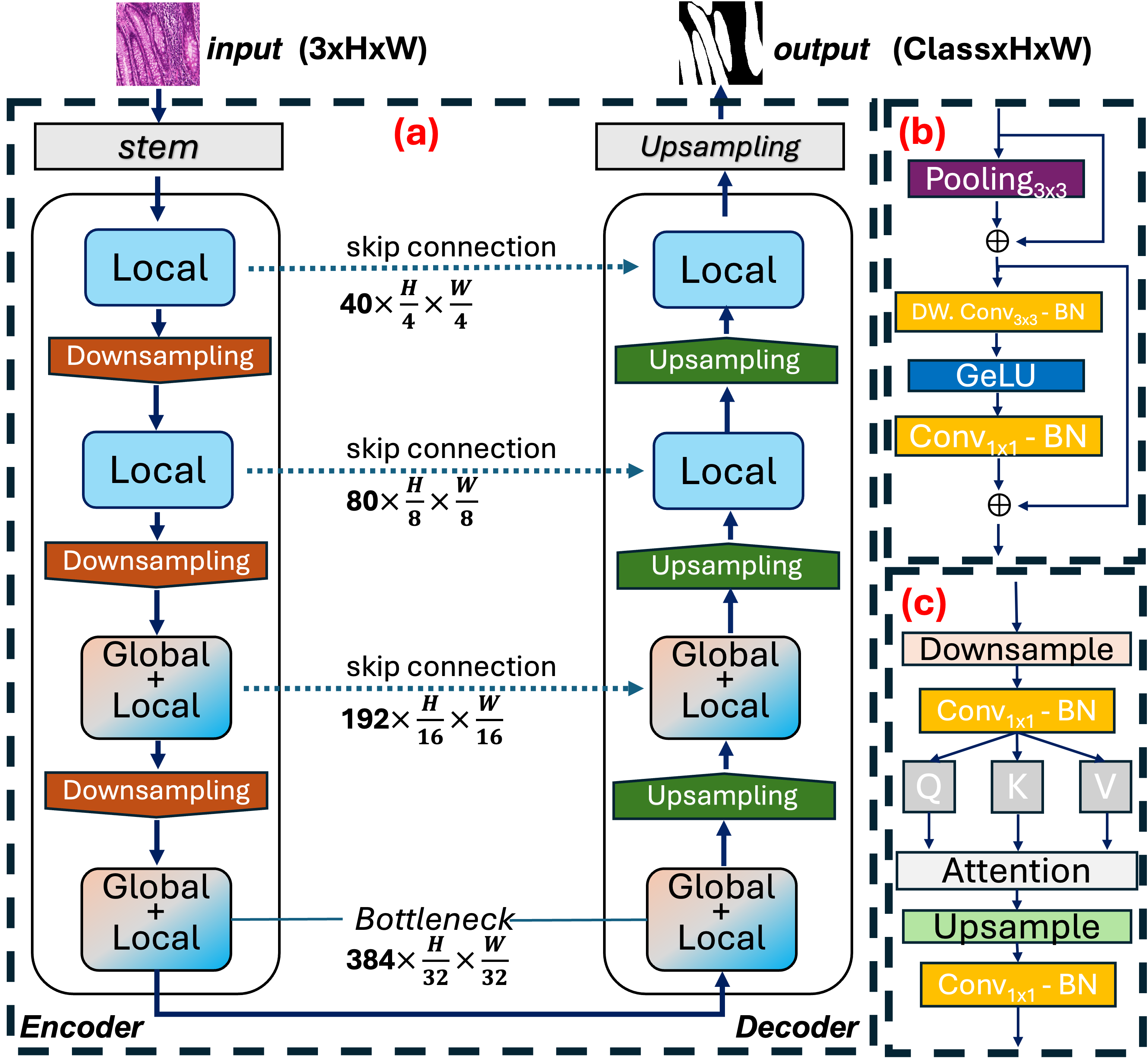}}
\caption{(a) The architecture of EViT-Unet, which is composed of encoder, bottleneck, decoder, and skip connections. "Local" blocks use convolution, while "Global+Local" blocks use the combination of convolution and self-attention. (b) The details of "Local" blocks. "DW.Conv," is the depthwise convolution~\cite{mobilenets}. (c) The details of "Global+Local" blocks.}

\label{fig:overview}
\end{figure}

\begin{table*}[!t]
\centering
\caption{Comparison with other methods on Synapse multi-organ CT dataset. The DSC denote dice similarity coefficient.}
\vspace{0.2cm}

\label{tab:synapse}
\resizebox{\textwidth}{!}{%
\begin{tabular}{l|c|cccccccc}
\toprule
\textbf{Methods} & \textbf{Average DSC} &  \textbf{Aorta} & \textbf{Gallbladder} & \textbf{Kidney(L)} & \textbf{Kidney(R)} & \textbf{Liver} & \textbf{Pancreas} & \textbf{Spleen} & \textbf{Stomach} \\ 
\hline
R50 U-Net ~\cite{U-net}& 74.68 & 87.74 & 63.66 & 80.60 & 78.19 & 93.74 & 56.90 & 85.87 & 74.16 \\
U-Net~\cite{U-net}& 76.85 & 89.07 & 69.72 & 77.77 & 68.60 & 93.43 & 53.98 & 86.67 & 75.58 \\
R50 Att-UNet~\cite{att-Unet}  & 75.57 & 55.92 & 63.91 & 79.20 & 72.71 & 93.56 & 49.37 & 87.19 & 74.95 \\
Att-UNet~\cite{att-Unet} & 77.77 & 89.55 & 68.88 & 77.98 & 71.11 & 93.57 & 58.04 & 87.30 & 75.75 \\
TransUNet~\cite{transunet}  & 77.48 & 87.23 & 63.13 & 81.87 & 77.02 & 94.08 & 55.86 & 85.08 & 75.62 \\
SwinUNet~\cite{swin-Unet}  & 79.13  & 85.47 & 66.53 & 83.28 & 79.61 & 94.29 & 56.58 & 90.66 & 76.60 \\
LeViT-UNet~\cite{levit-unet} & 78.53  & 78.53 & 62.23 & 84.61 & 80.25 & 93.11 & 59.07 & 88.86 & 72.76 \\
DeepLabv3~\cite{deeplabv3} & 77.63  & 88.04 & 66.51 & 82.76 & 74.21 & 91.23 & 58.32 & 87.43 & 73.53 \\
HiFormer~\cite{hiformer}  & 80.29 & 85.63 & 73.29 & 82.39 & 64.84 & 94.22 & 60.84 & 91.03 & 78.07 \\

SelfReg + UNet~\cite{selfreg,U-net} & 80.34& 88.74 & 71.78 & 85.32 & 80.71 & 93.80 & 62.22 & 84.78 & 75.39  \\ 
SelfReg + SwinUNet~\cite{selfreg,swin-Unet} & 80.54 & 86.07 & 69.65 & 85.12 & 82.58 & 94.18 & 61.08 & 87.42 & 78.22 \\

\hline
Ours & \textbf{80.87} & \text 87.13 & 66.53 & 85.45 & 83.14 & 94.92 & 62.92 & 89.66 & 77.18 \\

\toprule
\end{tabular}%
}

\end{table*}

\section{Method}
\label{sec:method}
\subsection{Architecture overview}

The overall architecture of our network is illustrated in Fig.~\ref{fig:overview}(a). The design consists of an encoder, decoder, bottleneck layer, and skip connections during the upsampling. Both the encoder and decoder are structured into four stages, and we employ EfficientFormerV2 block~\cite{efficientFormerV2} as the basic unit. The input through initial feature extraction block(stem) and downsampling to the size tp ${40 \times \frac{H}{4} \times \frac{W}{4}}$. Then, the input is downsampled after the block in each stage of the encoder, with downsampling rates of $2$. The encoder adopts the channel configuration shown in Fig.~\ref{fig:overview}(a). Global feature fusion occurs in the encoder’s final stage, and the features are passed to the decoder. We designed a decoder that is symmetric to the encoder. The decoder features are combined with the encoder features through skip connections, restoring the image features, and then performing 2x upsampling in each stage. Finally, the upsampling module performs $4x$ upsampling and outputs pixel-level predictions.

\subsection{Efficientformer block}
Different from models that purely use Transformer and self-attention as encoders, our network adopts a hybrid approach, combining convolution with self-attention modules. In the high-resolution stages, where self-attention requires calculating interactions between all pixels, leading to significant computational overhead~\cite{attention}, our blocks employ the depthwise(DW) convolution~\cite{mobilenets} to construct its feed-forward network (FFN) to extract local features, as illustrated in Fig~\ref{fig:overview}(b). Compared to standard convolution, DW convolution applies one filter per input channel, significantly reducing the computational complexity and enhancing the local features. The process can be described as:

\[
\mathbf{X}_{i+1,j} = S_{i,j} \cdot FFN\left( \mathbf{X}_{i,j} \right) + \mathbf{X}_{i,j}
\]
where the $i$ is the $i$th layer in the $j$ stage, And the $S$ is a learnable layer scale~\cite{metaformer}.


In the low-resolution stages, the computational burden of the self-attention mechanism is significantly reduced. Our blocks introduce the multi-head self-attention (MHSA) mechanism~\cite{attention}(Fig.~\ref{fig:overview}(c)), which enhances the ability to capture global features and enriches multi-scale features in the encoder. In the decoder, the multi-head attention mechanism improves the accuracy of image reconstruction by aggregating global and local features. This approach effectively balances accuracy and computational efficiency, allowing the model to capture complex global dependencies without significantly increasing the computational load. This process can be described as follows:


\[
\textit{X}_{i+1,j} = S_{i,j} \cdot MHSA\left(Proj\left(\textit{X}_{i,j} \right)\right) + \textit{X}_{i,j}
\]
The input features are projected through the mapping function \textit{Proj}$(\textit{X}_{i,j})$  via linear transformations to obtain the query (Q), key (K), and value (V) in the attention mechanism:

\[
\textit{MHSA}(Q,K,V) = \text{Softmax}(Q \cdot K^T + ab) \cdot V
\]
where ${ab}$ is the learnable attention bias for position encoding.

\subsection{Downsampling and Upsampling}


During the downsampling process, convolution is similarly employed in the high-resolution stages for efficient downsampling and to reduce the size of feature maps. In the low-resolution stages, we adopt the self-attention mechanisms for downsampling, which adjusts the number of query tokens to effectively capture global dependencies and multi-scale features during downsampling. This approach balances the computational complexity of the self-attention mechanism due to the reduced resolution. We propose a symmetric design for the decoder, utilizing self-attention upsampling in the low-resolution stages by adjusting the number of query tokens. At the same time, convolution operations are used to upsample the high-resolution stages. While ensuring accurate image reconstruction, it reduces computational complexity. The self-attention downsampling/upsampling can be described as:

\[
\textit{Out}_{[B,H,r \times N,C]} = \left( Q_{[B,H,r \times N,C]} \cdot K^T_{[B,H,C,N]} \right) \cdot V_{[B,H,N,C]}
\]

where ${r}$ is the scaling factor, When  ${r = \frac{1}{2}}$ , it represents downsampling, when ${r}$ = 2, it represents upsampling.

\subsection{Skip Connection}

In U-shaped segmentation networks, skip connections play a crucial role by passing the features collected from the encoder to the decoder to help retain low-level features effectively~\cite{U-net}. However, recent studies found some limitations of traditional skip connections~\cite{uctransnet}. Simply concatenating the encoder and decoder features may introduce redundancy, and since skip connections primarily pass local features, they struggle to capture global dependencies in more complex segmentation tasks. However, some studies introduced attention mechanisms into skip connections and have achieved promising results~\cite{uctransnet, att-Unet}. Based on the investigation and research, we introduced channel attention~\cite{uctransnet} into skip connections. This method enhances feature fusion by applying attention to emphasize important feature channels and suppress redundancy. It also facilitates better global dependency modeling across different feature scales, all while introducing minimal additional computational overhead. The channel-based skip connection can be described as follows:


\[
\text{att}_{avg} = \frac{\text{MLP}_x(\text{AVGPool}(x)) + \text{MLP}_g(\text{AVGPool}(g))}{2}
\]
\[
\text{out} = \text{ReLU}(x \times \textit{Sigmod}(\text{att}_{avg}))
\]
Here, $x$ is the current feature map from the previous layer, and $g$ is the skip connection feature map from the encoder during the downsampling process.

\begin{table}[htb]
\vspace{-0.2cm}
    \centering
    \hfill
    \begin{minipage}{1\columnwidth}
    \caption{Comparison of different methods in Glas and MoNuSeg datasets.}
    \vspace{0.2cm}
    \label{tab:glas&monuseg}
        \centering
        \resizebox{1\columnwidth}{!}{
        \begin{tabular}{l|cc|cc}
\hline
\multicolumn{1}{c}{{\textbf{Method}}} & \multicolumn{2}{c}{\textbf{Glas}} & \multicolumn{2}{c}{\textbf{MoNuSeg}} \\
\multicolumn{1}{c}{} & \multicolumn{1}{c}{ \textbf{DSC (\%)}} &\multicolumn{1}{c}{ \textbf{IOU (\%)}}& \multicolumn{1}{c}{ \textbf{DSC (\%)}} &\multicolumn{1}{c}{\textbf{ IOU (\%)}}  \\
\hline

U-Net~\cite{U-net}  & 85.45$\pm$1.25 & 74.78$\pm$1.67 & 76.45$\pm$2.62 & 62.86$\pm$3.00 \\
UNet++~\cite{unet++}  & 87.56$\pm$1.17 & 79.13$\pm$1.70 & 77.01$\pm$2.10 & 63.04$\pm$2.54 \\
AttUNet~\cite{att-Unet} & 88.80$\pm$1.07 & 80.69$\pm$1.66 & 76.67$\pm$1.06 & 63.47$\pm$1.16 \\
MRUNet~\cite{mru} & 88.73$\pm$1.17 & 80.89$\pm$1.67 & 78.22$\pm$2.47 & 64.83$\pm$2.87 \\
TransUNet~\cite{transunet} & 88.40$\pm$0.74 & 80.40$\pm$1.04 & 78.53$\pm$1.06 & 65.05$\pm$1.28 \\
MedT~\cite{medt} & 85.92$\pm$2.93 & 75.47$\pm$3.46 & 77.46$\pm$2.38 & 63.37$\pm$3.11 \\
SwimUNet~\cite{swin-Unet} & 89.58$\pm$0.57 & 82.06$\pm$0.73 & 77.69$\pm$0.94 & 63.77$\pm$1.15 \\
UCTransNet~\cite{uctransnet} & 90.18$\pm$0.71  & 82.96$\pm$1.06 & 79.08$\pm$0.67 & 65.50$\pm$0.91 \\
SelfReg + SwinUNet~\cite{selfreg} & 91.62$\pm$0.16  & 85.29$\pm$0.30  & 79.38$\pm$0.15 & 65.87$\pm$0.2 \\

\hline
Ours & \textbf{92.44$\pm$0.23}  & \textbf{86.50$\pm$0.38}  & \underline{79.27$\pm$0.24} & \textbf{65.87$\pm$0.21} \\

\hline
\end{tabular}
}
    \end{minipage}
\vspace{-0.2cm}
\end{table}

\begin{figure*}[!t]
    \centering
    \includegraphics[width=0.9\textwidth]{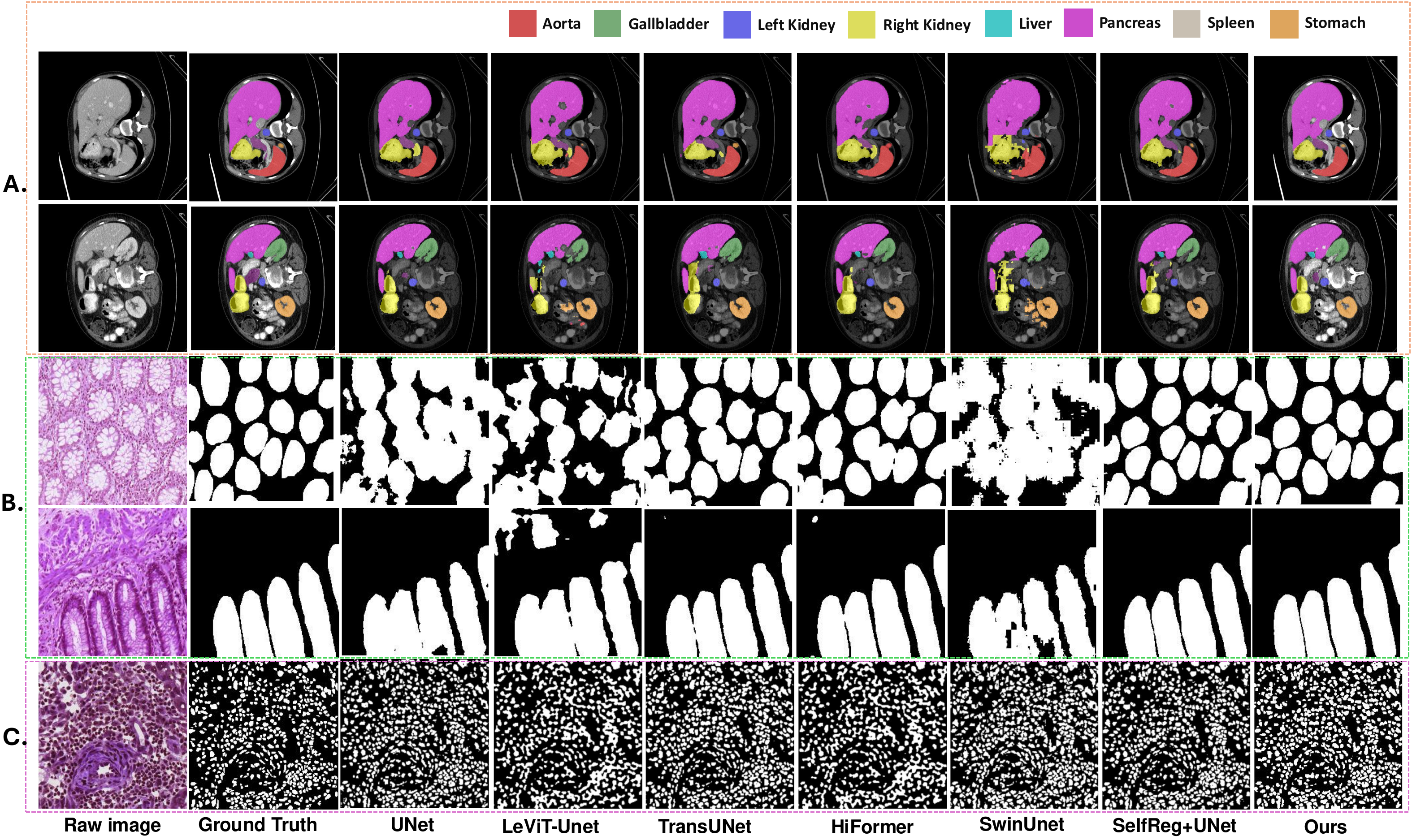}
    \caption{Comparison of segmentation results in Synapse(A), Glas(B) and MoNuSeg(C) dataset.}
    \label{fig:vis}
\end{figure*}

\section{Experiments And Results}
\label{sec:Experiments&Results}

\noindent \textbf{Synapse multi-organ segmentation dataset (Synapse)}~\cite{synapse} consists of 30 cases, with a total of 3779 axial abdominal clinical CT images. The dataset is split into 18 training and 12 testing samples. Our method is evaluated on eight abdominal organs, including the aorta, gallbladder, spleen, left kidney, right kidney, liver, pancreas, and stomach.

\noindent \textbf{Gland segmentation dataset (GlaS)~\cite{GlaS} and Multi-Organ Nucleus Segmentation (MoNuSeg)~\cite{MoNuSeg}} contains 85 images for training, 80 for testing, and 30 images for training and 14 for testing, respectively. We perform 5-fold cross-validation on the GlaS and MoNuSeg datasets. 

\noindent \textbf{Experiments Implement}
Our network is implemented based on Python 3.10 and PyTorch 2.0. Input images with size of 224x224, with a batch size of 32, and training is conducted on a single Nvidia A100 GPU. We employed the pre-trained weights of EfficientFormerV2 from ImageNet, adapting and loading them into both the encoder and decoder to initialize the model parameters as much as possible. Optimization is performed using the SGD optimizer with backpropagation.

\noindent \textbf{Results} The results are shown in Table~\ref{tab:synapse} for Synapse dataset, and Table~\ref{tab:glas&monuseg} for Glas and MoNuSeg datasets. Our model outperforms many popular current methods, achieving superior performance with 80.87\%. Specifically, it surpasses the best performance in our comparison method by 0.33\% in average DSC in the Synapse dataset. The results on the Glas and MoNuSeg datasets show that our method performs well on both datasets. Specifically, on the Glas dataset, our model achieved the best DSC of 92.44\% and an IOU of 86.50\%. On the MoNuSeg dataset, our model also achieved a DSC of 79.27\% and an IOU of 65.87\%, outperforming many popular comparison methods. We also obtain the visualization results for the Synapse dataset(Fig.~\ref{fig:vis}(A)), Glas dataset(Fig.~\ref{fig:vis}(B)), and MoNuSeg dataset(Fig.~\ref{fig:vis}(C)) to illustrate the performance of our method. Most importantly, we conducted a comparison of computational complexity on the Synapse dataset; our method outperforms all compared approaches in terms of computational efficiency, and the computational complexity is only \textbf{6.39 GMac}(Fig.~\ref{fig:Flops}).

\section{Conclusion And Discussion}
\label{sec:Conclusion}
In conclusion, we have developed a segmentation framework that achieves outstanding performance and offers superior computational efficiency. Our model surpasses numerous state-of-the-art methods in accuracy while maintaining a lower computational burden, making it particularly suited for limited computational resources devices, such as medical devices. These qualities emphasize the model’s suitability for performance-critical, real-world applications.

Despite the model’s outstanding performance and high efficiency, there are still some limitations regarding adaptability and practicality for medical devices. Medical devices have complex requirements, and while our framework performs well in experiments, further optimization is needed for broader deployment in embedded and portable systems. Looking ahead, our research can focus on fine-tuning the model for specific hardware implementations, thereby enhancing its applicability in real-world medical devices.

\section{ COMPLIANCE WITH ETHICAL STANDARDS}
This research study was conducted retrospectively using human subject data available in open access by Synapse~\cite{synapse}, GlaS~\cite{GlaS} and MoNuSeg~\cite{MoNuSeg}. Ethical approval was not required as confirmed by the license attached with the open-access data.
\section{ACKNOWLEDGMENTS}
This work was supported by grants from the National Institutes of Health (R01EY032125 and R01DE030286) and the State of Arizona via the Arizona Alzheimer Consortium.


\bibliographystyle{IEEEbib}
\bibliography{main}

\end{document}